\documentclass[a4paper,10pt,twoside]{cpc-hepnp}
\usepackage{CJK,upgreek,fancyhdr}
\usepackage{multicol}
\usepackage{graphicx}
\usepackage{booktabs}
\usepackage{amssymb,bm,mathrsfs,bbm,amscd}
\usepackage[tbtags]{amsmath}
\usepackage{lastpage}
\usepackage{subfigure}
\usepackage[justification=centering]{caption}
\usepackage{textcomp}

\begin{document}
\begin{CJK*}{GB}{gbsn}

\fancyhead[c]{\small Chinese Physics C~~~Vol. xx, No. x (201x) xxxxxx}
\fancyfoot[C]{\small 010201-\thepage}

\footnotetext[0]{Received 31 June 2015}

\title{A hybrid structure gaseous detector for ion backflow suppression\thanks{Supported by National Key Programme for S\&T Research and Development (2016YFA0400400)
and by National Natural Science Foundation of China (11275224) }}

\author{%
      Yu-Lian Zhang(ÕÅÓàÁ¶)$^{1,2,3}$%
\quad Hui-Rong Qi(Æî»ÔÈÙ)$^{2,3;1)}$\email{qihr@ihep.ac.cn}
\quad Bi-Tao Hu(ºú±ÌÌÎ)$^{1;2)}$\email{hubt@lzu.edu.cn}
\quad Hai-Yun Wang(Íõº£ÔÆ)$^{2,3,4}$
\\Qun Ou-Yang(Å·ÑôȺ)$^{2,3}$
\quad Yuan-Bo Chen(³ÂÔª°Ø)$^{2,3}$
\quad Jian Zhang(ÕŽ¨)$^{2,3}$
\quad Zhi-Wen Wen(ÎÂÖ¾ÎÄ)$^{1,2,3}$
}
\maketitle

\address{%
$^1$ School of Nuclear Science and Technology, Lanzhou University, Lanzhou 730000, China\\
$^2$ State Key Laboratory of Particle Detection and Electronics, Beijing 100049, China\\
$^3$ Institute of High Energy Physics, Chinese Academy of Sciences, Beijing 100049, China\\
$^4$ Graduate University of Chinese Academy of Sciences, Beijing 100049, China\\
}

\begin{abstract}
  A new concept for a hybrid structure gaseous detector module with
  ion backflow suppression for the time projection chamber in a future circular collider is presented.
  It is a hybrid structure cascaded Gas Electron Multiplier (GEM) with a Micromegas detector.
  Both Micromegas and GEM have the capability to naturally reduce most of the ions produced in the amplification region.
  The GEM also acts as the preamplifer device and increases gas gain together with the Micromegas.
  Feasibility tests of the hybrid detector are performed using an $^{55}$Fe X-ray source.
  The energy resolution is better than 27$\%$ for 5.9\,keV X-rays.
  It is demonstrated that a backflow ratio better than 0.2$\%$ can be reached in the hybrid readout structure at a gain of 5000.
\end{abstract}



\footnotetext[0]{\hspace*{-3mm}\raisebox{0.3ex}{$\scriptstyle\copyright$}2013
Chinese Physical Society and the Institute of High Energy Physics
of the Chinese Academy of Sciences and the Institute
of Modern Physics of the Chinese Academy of Sciences and IOP Publishing Ltd}%

\begin{multicols}{2}

\section{Introduction}

Time Projection Chambers (TPCs) have been extensively studied and used in many fields, especially in particle physics experiments, including STAR~[1] and ALICE~[2].
Their low material budget and excellent pattern recognition capability make them ideal for three-dimensional tracking and identification of charged particles.
They are also the only type of electronically read gaseous detector delivering direct three-dimensional track information~[3].
However, there has always been a critical problem with TPCs, especially in high background conditions -- the space charge distortion due to the accumulation of positive ions in the drift volume.
Due to their large mass, positive ions move slowly under the action of electric field in the drift volume of the TPC.
The continuously superimposed ions in the drift volume of the TPC may affect the drift behaviour of electrons from a later track~[4].
The majority of ions inside the drift volume are backflowing ions from the  amplification region of the TPC readout devices.
It is thus of great importance to limit ion backflow (IBF) from the amplification region.

Early TPCs were equipped with multi-wire proportional chambers (MWPCs)~[5] as gas amplification devices.
The IBF ratio in a standard MWPC is 30-40{\%}~[6], so a gating grid is essential to prevent ions from reaching the drift volume~[7].
In the presence of a trigger, the gating grid switches to the open state to allow ionization electrons to travel into the gas amplification region.
After a maximum drift time of about 100\,\textmu s (depending on the drift length, electric field and gas mixture), the gating grid is closed to prevent positive ions from drifting back into the drift volume.
Since it must remain closed until the ions have been collected on the grid wires, the ionization electrons are also blocked during this time and
the dead time consequently increases.
Triggered operation of a gating grid will therefore lead to loss of data. 
Thus, the TPC at the proposed circular collider will have to be operated continuously and the backflow of ions must be minimized without the use of a gating grid.

TPC readout with micro-pattern gaseous detectors (MPGDs), especially Gas Electron Multipliers (GEM)[8] and micro-mesh gaseous structures (Micromegas)[9], is very attractive,
because the IBF of those detectors is intrinsically low, usually around a few percent.
GEM detectors have been extensively proved in the last decade to be the prime candidate, as they offer excellent results for spatial resolution and low IBF~[10,11].
Several GEM foils can be cascaded, allowing multilayer GEM detectors to be operated at an overall gas gain above 10$^{4}$  in the presence of highly ionized particles.
Micromegas is another kind of MPGD that is likely to be used as end-cap detectors for the TPC readout.
It is a parallel plate device, composed of a very thin metallic micromesh which separates the detector region into drift and amplification volumes.
The IBF of this detector is equal to the inverse of the field ratio between the amplification and the drift electric fields~[12].
Low IBF therefore favours high gain.
However, high gain will make it particularly vulnerable to sparking~[13].
The idea of combining GEM with Micromegas was first proposed with the goal of reducing the spark rate of Micromegas detectors~[14].
Preamplification using GEMa also extends the maximum achievable gain, so there have also been studies on gaseous photomultipliers with this hybrid configuration~[15].

To fulfill the physics goals of the future circular collider, a TPC with excellent performance is required.
MPGDs with outstanding single-point accuracy and excellent multi-track resolution are needed.
We have proposed and investigated the performance of a novel configuration detector module: a combination of GEM and a Micromegas.
The detector will be called GEM-MM for short throughout this paper.
The aim of this study is to suppress IBF continually by eliminating the gating grid.
The design concept and some results of the prototype module are described in the paper.


\section{Experimental setup}

The cascaded structure of the GEM-MM detector is composed of a drift electrode, a GEM foil, a standard Micromegas, and a readout printed circuit board.
The Micromegas detector is based on the bulk method and has an active area of 25\,mm¡Á25\,mm.
The micromesh is made of stainless steel wires 22\,\textmu m in diameter, interwoven at a pitch of 62\,\textmu m.
128\,\textmu m under the micromesh is a single copper pad readout plane.
A GEM foil is cascaded above the micromesh at a distance of 1.4\,mm.
It is a standard GEM foil of area 25\,mm¡Á25\,mm, obtained from CERN.
In the experiment, the drift distance was maintained at 4\,mm.
Electrodes were biased with CAEN N471A high voltage units.
A $^{55}$Fe source was used to produce the primary electrons in the sensitive volume during the test.
The working gas was a mixture of Ar/CO$_{2}$(90/10) at room temperature and atmospheric pressure.

In the gain measurement (shown schematically in Fig. 1), the micromesh  electrode was connected to a charge sensitive pre-amplifier (ORTEC model 142IH) and the anode pad was grounded.
Subsequently, the output pulse from the pre-amplifier was fed to an amplifier (ORTEC model 572A) with a shaping time constant of 1\,\textmu s.
The data was finally recorded in a multi-channel analyzer (ORTEC ASPEC 927).
To characterize the performance of the GEM-MM detector, the electronic gain was calibrated in advance.
Under various GEM and micromesh voltages, $^{55}$Fe spectra were then taken.
The detector gains are available from the spectra with the calibrating result of electronic gain.

Ion backflow is due to secondary ions generated in an electron-avalanche process in the amplification which return to the drift space.
In this paper, fractional ion feedback is defined as the ratio of the ion charge injected into the drift volume, collected on the drift electrode, and the electron charge collected on the anode pad.
In the experimental test of IBF, the currents on the drift cathode and the anode pad are measured as $\mathrm{I_{c}}$ and $\mathrm{I_{a}}$ respectively.
$\mathrm{I_{c}}$ is proportional to the number of ions collected on the drift cathode.
Positive charge collected on the drift cathode is primarily backflowing ions from the gas amplification region (avalanche region and GEM hole as shown in Fig. 1). The contribution from ions created during the primary ionization process is also included.
It is not negligible since the backflowing ion current is very small with a $^{55}$Fe source.
The primary ion current $\mathrm{I_{prim}}$ of the drift cathode is also measured, as shown in the next section.
$\mathrm{I_{a}}$ is the current measured in the anode pad and is proportional to the number of electrons collected on the anode.
The formula for the IBF is:
\begin{equation}
\label{two}
\mathrm{IBF} = \frac{\mathrm{I_{c}}-\mathrm{I_{prim}}}{\mathrm{{I_{a}}}}
\end{equation}
A Keithley (6517B~[16])  pico current electrometer  was used to measure the current with a resolution of $\sim$1\,pA.
The electrode to be measured was at a grounded potential. 
In the measurement of anode current, the detector was operated with a negative voltage. The further the electrode is from the anode, the more negative the bias potential.
For the measurement of current from the drift cathode's ions,
the drift cathode was grounded, and the other electrodes biased with positive voltages to maintain the electric field as for the anode current measurements.


\section{Results and discussion}

\subsection{Energy spectrum}
A $^{55}$Fe X-ray source with a characteristic energy of 5.9\,keV was used in the test.
In the argon-based working gas mixture, a typically pulse height spectrum for a GEM or Micromegas detector contains one major peak corresponding to the 5.9\,keV X-rays and an escape peak at lower pulse heights corresponds to the ionization energy of an electron from the argon K-shell.
In the GEM-MM detector, the situation is different.
There are two amplification stages inside this detector.
The primary ionization created by photon absorption can be in the drift region or in the transfer region (Fig. 1).
Photoelectrons starting from the drift region get amplified by both the GEM detector and the Micromegas detector before they are collected on the anode.
If the photons are absorbed in the transfer region, the primary electrons will be amplified only once (by Micromegas).
    \end{multicols}
    \begin{figure}[!htbp]
    \begin{center}
    \includegraphics[width=13cm,height=4.5cm]{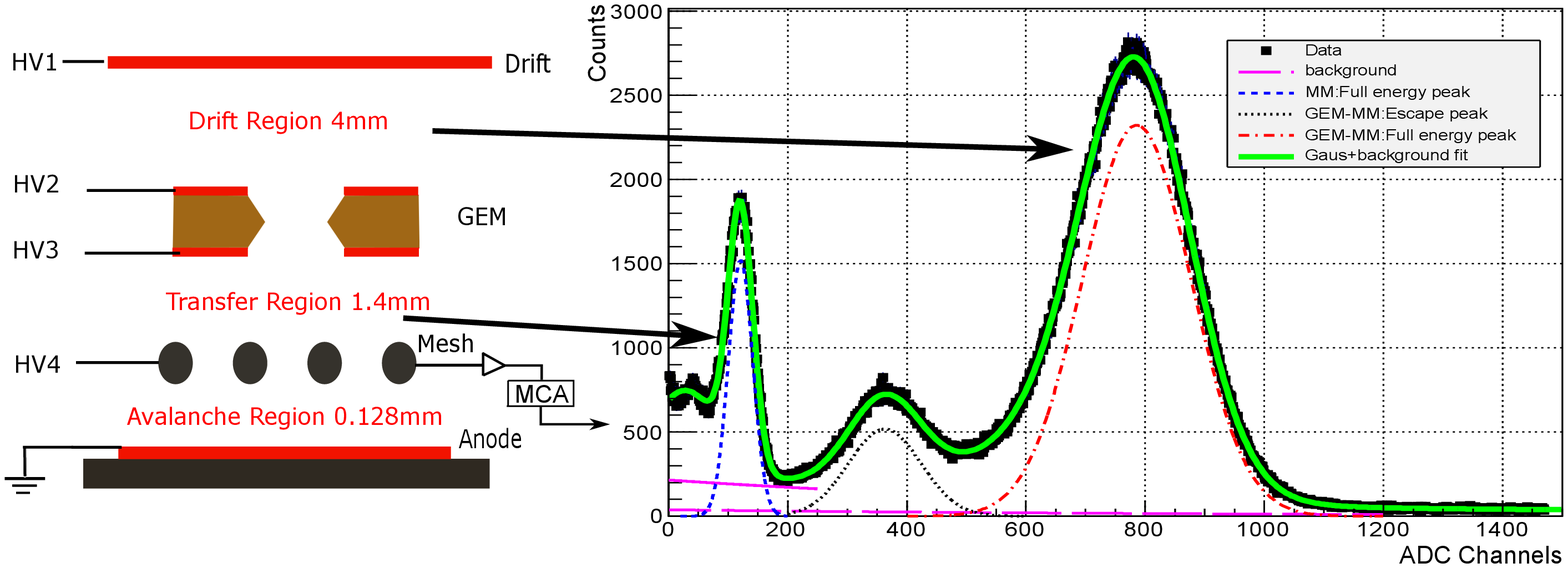}
    \figcaption{\label{fig1} (color online) GEM-MM configuration (left) and pulse height spectrum at 5.9 keV for a GEM-MM, showing each peak and the corresponding location of primary ionization (right). Source: $^{55}$Fe; gas: Ar(90)+CO$_{2}$(10); E$_{d}$=250\,V/cm, V$_{gem}$=340\,V, E$_{t}$=500\,V/cm, V$_{mesh}$=420\,V.}
    \end{center}
    \end{figure}
    \begin{multicols}{2}
Figure 1 depicts a typical $^{55}$Fe pulse height spectrum obtained by the GEM-MM detector.
Four peaks are seen in the pulse height spectrum. From left, the first peak and the second peak are the escape peak and the full energy peak of the standalone Micromegas.
The last two peaks are created by photons with their energy deposited in the drift region.
These primary electrons show combination amplification.
The principle of the GEM-MM detector is fully verified.

\begin{center}
\tabcaption{ \label{tab1}  Gaussian fit results.}
\footnotesize
\begin{tabular*}{80mm}{c@{\extracolsep{\fill}}ccc}
\toprule Peak & Mean   & Sigma  & Resolution({\%}) \\
\hline
MM Photo Peak\hphantom{00} & \hphantom{0}120.9 & \hphantom{0}20.6 & \hphantom{0} 40.1 \\
GEM-MM Escape Peak\hphantom{0} & \hphantom{0}362.9 & \hphantom{0}60.8 & \hphantom{0} 39.4 \\
GEM-MM Photo Peak\hphantom{0} & \hphantom{0}785.9 & \hphantom{0}91.1 & \hphantom{0} 27.3 \\
\bottomrule
\end{tabular*}
\end{center}

The spectrum is fitted with a Gaussian distribution to find the mean and the sigma for the four peaks.
Table~1 summarizes the Gaussian fit mean and sigma values for three of the four peaks.
The escape peak of the Micromegas detector is not fitted well due to the electronics noise.
The energy resolution of each peak (FWHM) is also presented in the table.

\subsection{Gain}

With the calibrated electronic gain results, the gain of the detector is obtained from the measured spectra as described in the previous subsection.
The gain of the Micromegas, $\mathrm{G_{MM}}$, is characterised by the first full energy peak in the spectrum (MM photo peak in Table~1). The last full energy peak (GEM-MM photo peak in Table 1) represents the overall gain of the GEM-MM detector, $\mathrm{G_{GEM-MM}}$.
So the effective GEM gain can be expressed as:

\begin{equation}
\label{two}
\mathrm{G_{GEM}} = \frac{\mathrm{G_{GEM-MM}}}{\mathrm{G_{MM}}}
\end{equation}

\end{multicols}
\begin{figure}[!htbp]

\centering
\subfigure[]{
\label{fig:2:a} 
\includegraphics[width=7cm,height=4.5cm]{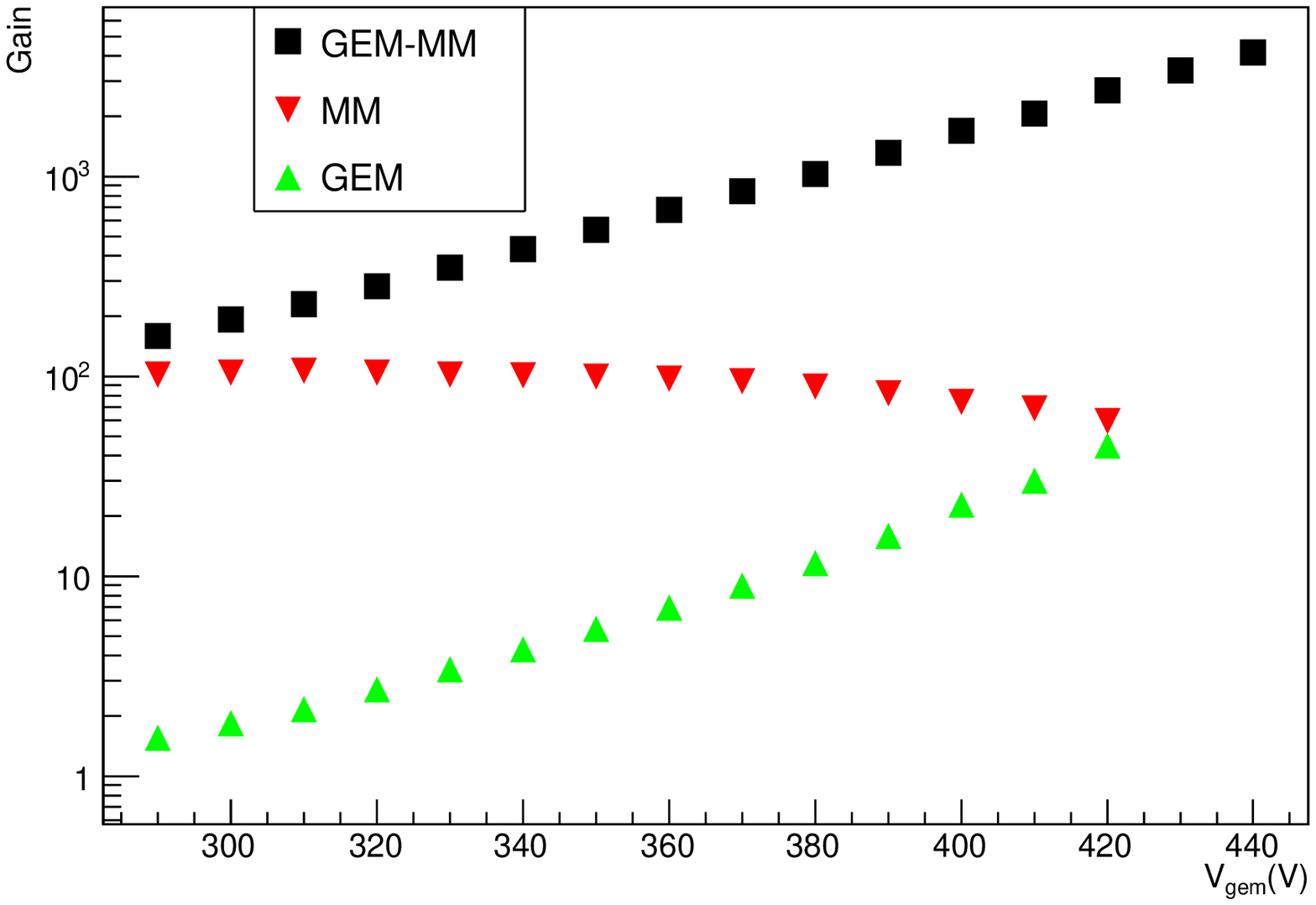}}
\hspace{0.01in}
\subfigure[]{
\label{fig:2:b} 
\includegraphics[width=7cm,height=4.5cm]{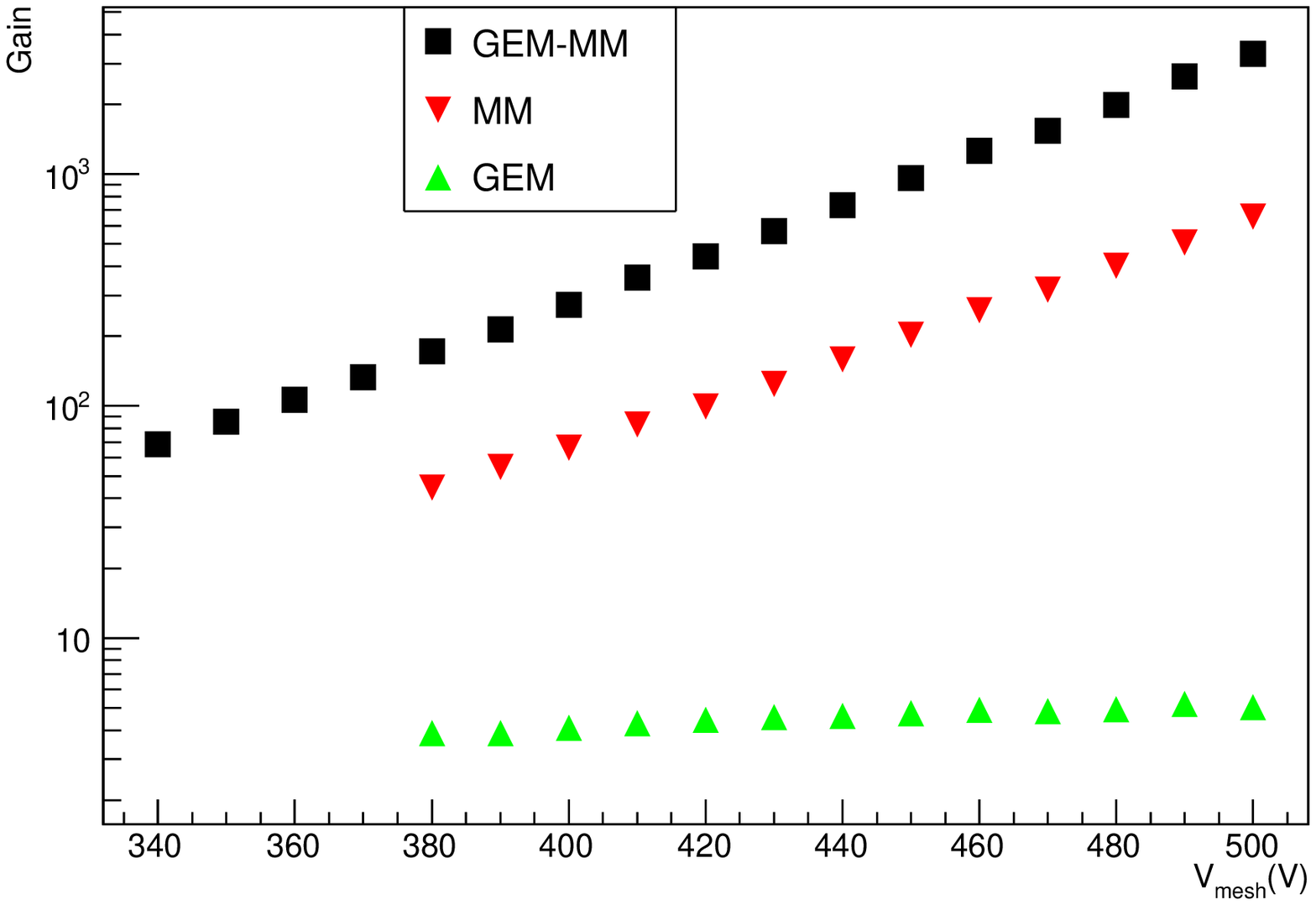}}
\hspace{0.1in}
\caption{ (color online) Gas gain versus GEM voltage, for micromesh V$_{mesh}$ = 420\,V (a) and micromesh voltage, V$_{gem}$ = 340\,V (b). E$_{d}$ = 250\,V/cm, E$_{t}$ = 500\,V/cm.}
\label{fig:2} 
\end{figure}
\begin{multicols}{2}
The GEM is the first gain element in the GEM-MM detector, and its effective gain is a function of the voltage across the GEM. Similarly, the gain of the Micromegas is a function of the voltage difference between the micromesh and the anode. Keeping the drift field at 250\,V/cm (typical drift field for TPC) and the transfer field at 500\,V/cm, a set of gain test results with various voltage settings are displayed in Fig. 2.
As shown in the figure, the GEM preamplification helps the GEM-MM achieve high gains with Micromegas working under relatively low voltages. The spark rates can be greatly reduced even at high overall gain.
It is important to note that this is a new way to measure the effective gas gain of a GEM.
A gain of 5000 or more can be achieved without any obvious discharge behaviour.

\subsection{Ion Backflow}
Another role of the GEM is to reduce the ion backflow.
With a precise measurement of the currents on the anode and the drift cathode it is possible to calculate the IBF, using Eq. (1), for different working conditions of the GEM-MM.
In Fig. 3, the correlations between voltage applied on the GEM (Fig. 3(a)) or the  Micromegas (Fig. 3(b)) and the currents measured on the anode or the drift cathode are demonstrated.
Figure 4 shows the calculated fractional IBF.
The gas gain of the GEM-MM detector is also plotted in the figures.

\end{multicols}
\begin{figure}[!htbp]

\centering
\subfigure[]{
\label{fig:3:a} 
\includegraphics[width=7cm,height=4.5cm]{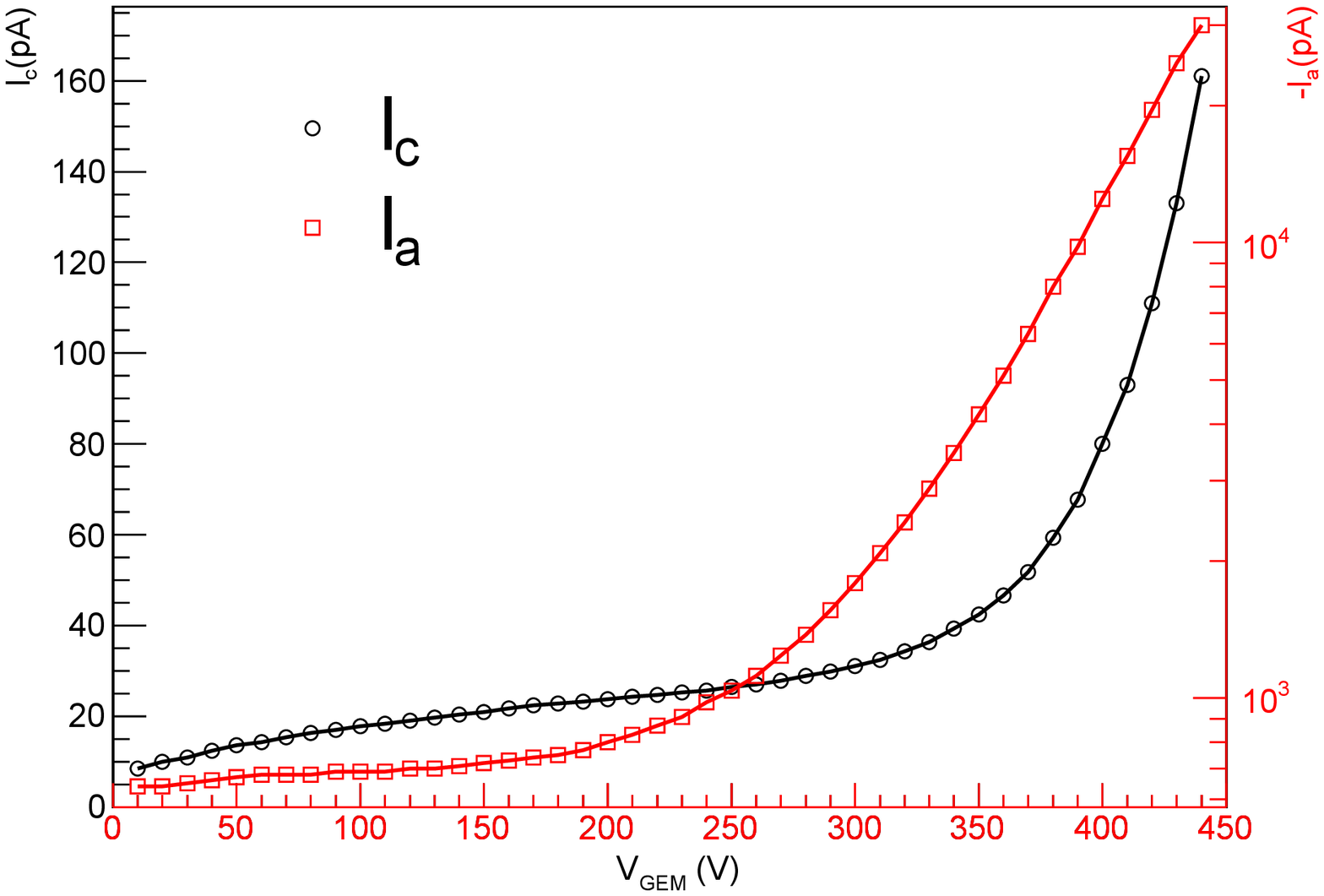}}
\hspace{0.01in}
\subfigure[]{
\label{fig:3:b} 
\includegraphics[width=7cm,height=4.5cm]{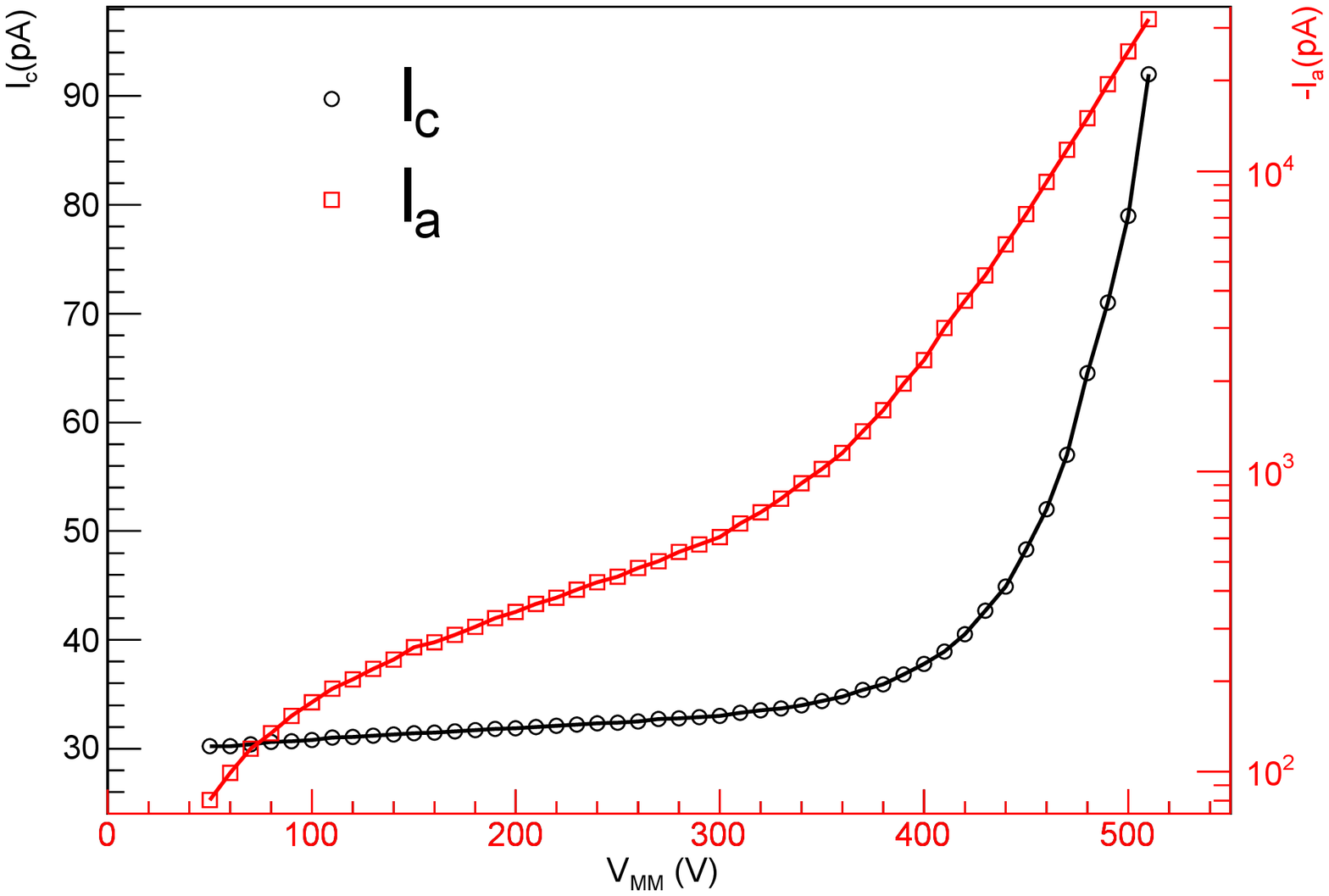}}
\hspace{0.1in}
\caption{ (color online) Measured currents on the anode (I$_{a}$) and the drift cathode (I$_{c}$) versus (a) GEM voltage, micromesh V$_{mesh}$ = 420\,V  and (b) micromesh voltage, V$_{gem}$ = 340\,V. E$_{d}$ = 250\,V/cm, E$_{t}$ = 500\,V/cm.}
\label{fig:3} 
\end{figure}
\begin{multicols}{2}

In Fig. 3(a) and Fig. 4(a), the voltage across the GEM foil begins at 10\,V and the Micromegas is working stably.
Except for primary ionization generated ions in the drift and transfer region, the ions collected on the drift cathode are from the avalanche region of the Micromegas detector.
With the increase of GEM voltage, the electron current on the anode remains the same before electron avalanche happens inside the GEM foil.
Nevertheless, the GEM foil will have an increasingly higher transparency for ions to pass through to the drift region, which means the ion current on the drift cathode increases.
Consequently, IBF increases as the GEM voltage increases.
However, gas amplification begins to occur inside the GEM hole as its voltage goes on increasing, which has a positive effect on the increase of the electron current on the anode.
Therefore, IBF increases initially and decreases afterwards as the GEM voltage increases.
So, an ion backflow value of {$\sim$3\%} is considered to be the IBF for a standalone Micromegas detector with a gain of about 600.
When the GEM is cascaded, the IBF can be further reduced to below {1\%}.
Figure 4(b) shows that when a constant bias voltage is set across the GEM, the  IBF decreases as the micromesh voltage increases.
The reason is that electrons collected on the anode increase with the increase of the mesh voltage.
So the IBF can be estimated as a few percent for a single GEM detector with a comparatively low gain of approximately 4. After the Micromegas is cascaded, however, IBF is reduced significantly.
An IBF ratio of 0.19$\%$ under overall detector gain of 5000 was achieved in our test. 
\end{multicols}
\begin{figure}[!htbp]

\centering
\subfigure[]{
\label{fig:7:a} 
\includegraphics[width=7cm,height=4.5cm]{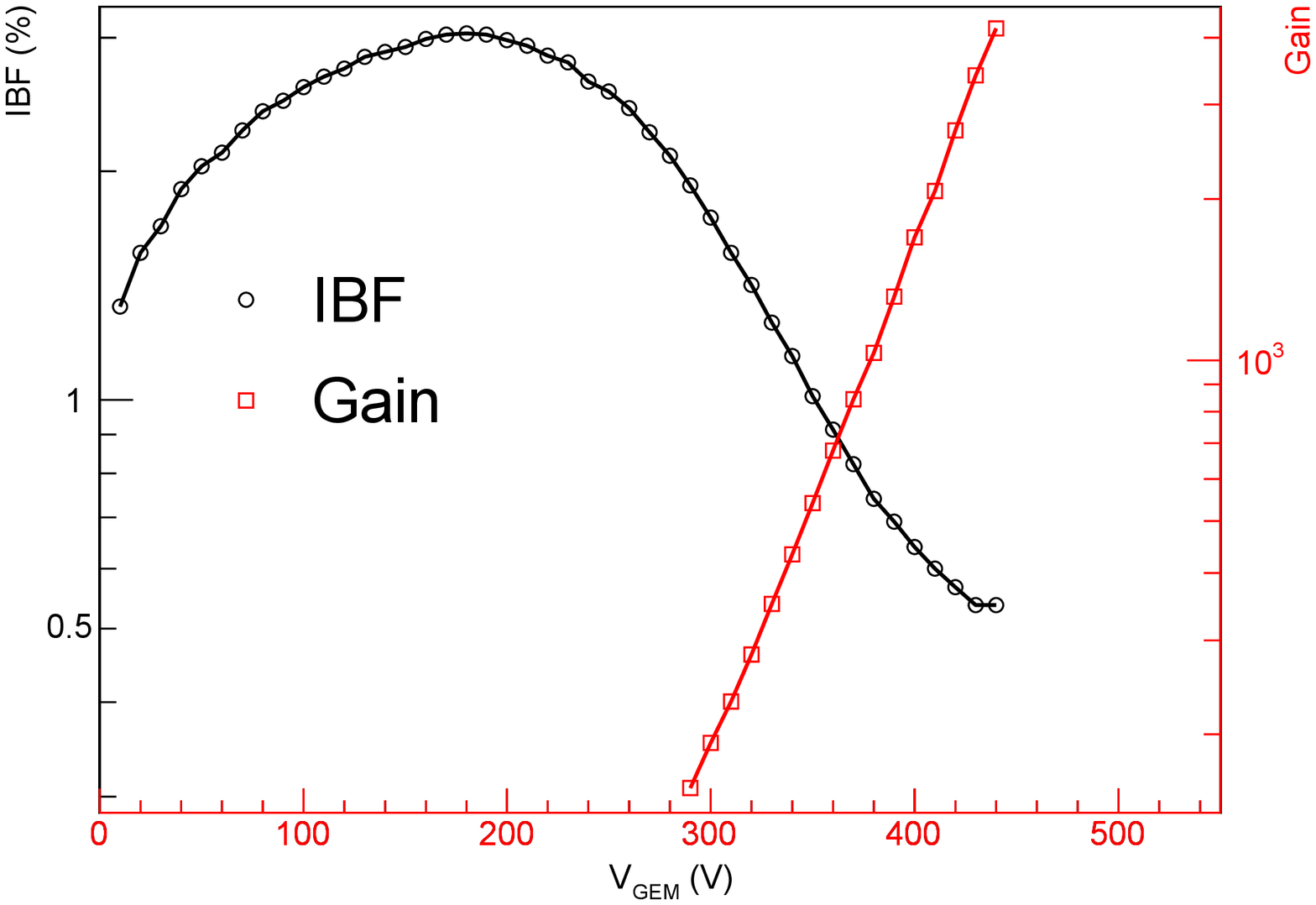}}
\hspace{0.01in}
\subfigure[]{
\label{fig:7:b} 
\includegraphics[width=7cm,height=4.5cm]{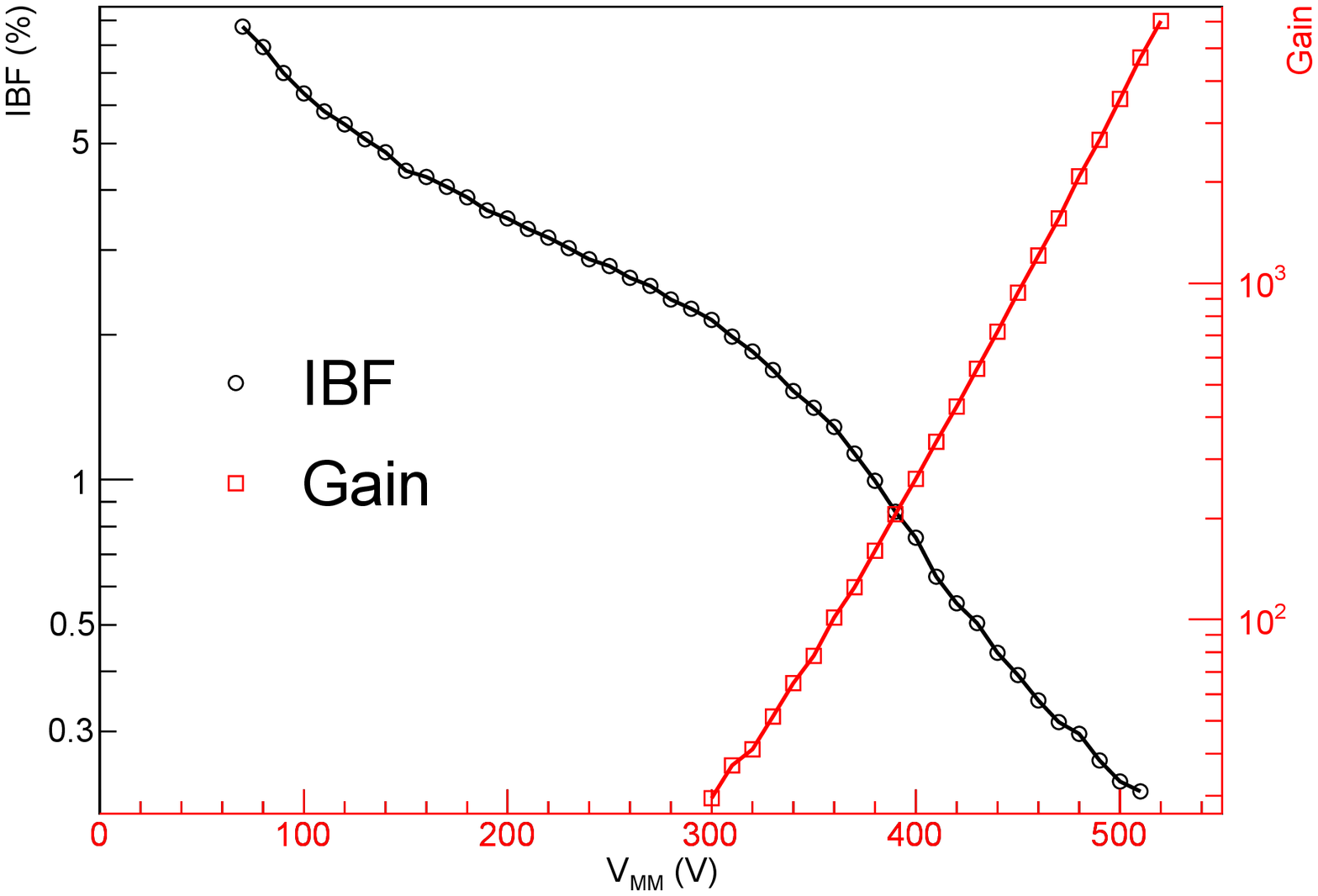}}
\hspace{0.1in}
\caption{ (color online) Gas gain and IBF versus (a) GEM voltage, micromesh V$_{mesh}$ = 420\,V  and (b) micromesh voltage, V$_{gem}$ = 340\,V. E$_{d}$ = 250\,V/cm, E$_{t}$ = 500\,V/cm.}
\label{fig:7} 
\end{figure}
\begin{multicols}{2}


\section{Conclusion}

In this paper, we have presented a new concept of avalanche ion backflow reduction for a future MPGD readout based TPC, and a  prototype has been developed.
It is a hybrid structure with one GEM foil cascaded above the Micromegas detector.
Tests of this detector have been carried out with an $^{55}$Fe X-ray source in Ar-CO${_2}$(90-10) gas mixture.
The preamplification effect of GEM foil has been demonstrated in the energy spectrum measurement.
With the novel hybrid structure, the effective gain of the GEM can be measured even when it is relatively low.
The energy resolution of this hybrid structure gaseous detector is measured to be 27\% (FWHM).
The gain properties of this device were measured. A gain up to about 5000 can be achieved without any obvious discharge behaviour.
The currents on the anode and drift cathode were measured precisely with an electrometer.
Out experimental measurements show that IBF can be reduced down to 0.19$\%$ at a gain of about 5000.
\\

\acknowledgments{The authors would like to thank Dr. XIA Xin for the useful discussions.}

\end{multicols}

\vspace{15mm}

\vspace{-1mm}
\centerline{\rule{80mm}{0.1pt}}
\vspace{2mm}

\begin{multicols}{2}

\end{multicols}

\clearpage
\end{CJK*}
\end{document}